\RequirePackage{lineno}
\documentclass[floatfix,twocolumn,showpacs,preprintnumbers,amsmath,amssymb,prl,superscriptaddress]{revtex4} 
\usepackage{url}
\usepackage{graphicx}
\usepackage{dcolumn}
\usepackage{bm}
\usepackage{soul}
\usepackage{ulem}

\usepackage{color}
\usepackage{natbib}
\begin{document}
\raggedbottom
\title{
Elliptic and quadrangular flow of protons in the high baryon density region}
\author{Shaowei Lan}\affiliation{School of Electrical and Mechanical Engineering, Pingdingshan University, 467000 Pingdingshan, China}
\author{Zuowen Liu}\affiliation{Key Laboratory of Quark and Lepton Physics (MOE) and Institute of Particle Physics, Central China Normal University, Wuhan 430079, China}
\author{Like Liu}\affiliation{Key Laboratory of Quark and Lepton Physics (MOE) and Institute of Particle Physics, Central China Normal University, Wuhan 430079, China}
\author{Shusu Shi}\affiliation{Key Laboratory of Quark and Lepton Physics (MOE) and Institute of Particle Physics, Central China Normal University, Wuhan 430079, China}

\begin{abstract}
The collective flow provides valuable insights into the anisotropic expansion of particles produced in heavy-ion collisions and is sensitive to the equation of the state of nuclear matter in high-baryon-density regions. In this paper, we use the hadronic transport model SMASH to investigate the elliptic flow ($v_2$), quadrangular flow ($v_4$), and their ratio ($v_{4}/v_{2}^{2}$) in Au+Au collisions at high baryon density. Our results show that the inclusion of baryonic mean-field potential in the model successfully reproduces experimental data from the HADES experiment, indicating that baryonic interactions play an important role in shaping anisotropic flow. In addition to comparing the transverse momentum ($p_T$), rapidity, and centrality dependence of $v_{4}/v_{2}^{2}$ between HADES data and model calculations, we also explore its time evolution and energy dependence across $\sqrt{s_{NN}} =$ 2.4 to 4.5 GeV. While the ratio $v_{4}/v_{2}^{2}$ for high-$p_{T}$ particles approaches 0.5, which aligns with expectations from hydrodynamic behavior, we emphasize that this result primarily reflects agreement with the HADES measurements rather than a definitive indication of ideal fluid behavior. These findings contribute to understanding the early-stage dynamics in heavy-ion collisions at high baryon density.

\end{abstract}

\maketitle
\realpagewiselinenumbers
\setlength\linenumbersep{0.10cm}

\section{Introduction}
\label{sec:intro}

The study of nuclear matter under extreme conditions is one of the fundamental goals in heavy-ion collision experiments, where the creation of the quark--gluon plasma (QGP), a state of deconfined quarks and gluons, offers insight into the early universe and the strong interaction described by Quantum Chromodynamics (QCD)~\cite{Shuryak:1978ij,Bzdak:2019pkr,Braun-Munzinger:2007edi}. 
The QGP is expected to form in environments with extremely high temperatures or near-zero baryon densities, and different regimes of QGP can be probed depending on the collision energy. At high collision energies, such as those achieved at the Large Hadron Collider (LHC) and at the top energy of the Relativistic Heavy-Ion Collider (RHIC), the QGP behaves like an almost perfect fluid, exhibiting strong collective flow patterns driven by pressure gradients in the system~\cite{STAR:2007mum,STAR:2015gge,ALICE:2011ab,ATLAS:2012at, Gale:2012rq, Schenke:2010rr, Schenke:2010nt, Alver:2010dn, Bozek:2011ua, Niemi:2011ix, Niemi:2015qia, Shen:2020jwv, Nara:2019qfd}.  
When the collision energy decreases, in the beam energy scan region at RHIC, the system provides access to a more complex phase diagram, where the phase transition from QGP to hadronic matter occurs~\cite{Hung:1994eq,STAR:2020tga,STAR:2014clz}. 
At a few GeV collision energy and high baryon density, for example, in the ongoing fixed-target program at the HADES experiment~\cite{HADES:2020lob}, the fixed-target program at STAR BES-II~\cite{STAR:2021yiu}, as well as at the upcoming Facility for Antiproton and Ion Research (FAIR)~\cite{CBM:2016kpk}, the nuclotron-based ion collider facility (NICA)~\cite{Kekelidze:2016wkp}, and the High-Intensity Heavy-Ion Accelerator Facility (HIAF)~\cite{Yang:2013yeb}, the system evolves under significantly different conditions. The presence of spectator nucleons, baryon stopping, and enhanced hadronic interactions in this regime can strongly affect the development of flow~\cite{STAR:2021yiu, Liu:2024ugr,HADES:2020lob, HADES:2022osk, Luo:2020pef, Ivanov:2024gkn, Shen:2020jwv, Nara:2019qfd, Oliinychenko:2022uvy, Steinheimer:2022gqb, OmanaKuttan:2022aml, Li:2022cfd, Wu:2023rui, Mohs:2024gyc}.

Anisotropic flow is a widely used tool for understanding the properties of hot and dense matter created in heavy-ion collisions. The flow observables are determined by the azimuthal anisotropy of emitted particles with respect to the reaction plane~\cite{Sorge:1996pc,Ollitrault:1993ba,Poskanzer:1998yz}. The azimuthal distribution is commonly expressed through a Fourier decomposition, as shown below:
\begin{equation}
    E\frac{d^{3}N}{d^{3}p} = \frac{1}{2\pi} 
    \frac{d^{2}N}{p_{T}dp_{T}dy}
    \left[ 
    1+2\sum_{n=1}^{\infty}v_{n}cos[n(\phi-\Psi_{RP})]
    \right]
    \label{eq:eq1}
\end{equation}
where $\phi$ is the azimuthal angle of the particles and $\Psi_{RP}$ is the reaction plane angle.  
The second- and fourth-order coefficients are called elliptic flow $v_2$ and quadrangular flow $v_4$. These observables provide valuable insights into the system's expansion dynamics, reflecting initial spatial asymmetries in momentum space during the evolution of the QGP or hadronic matter~\cite{Sorge:1996pc,Kolb:2003zi}.
Theoretical calculations suggest that the quadrangular flow divided by the squared elliptic flow can serve as a probe of the system's ideal hydrodynamic behavior when this ratio, $v_4/v_2^2$, is approximately 0.5~\cite{Borghini:2005kd}. 
Measurements from RHIC-PHENIX and RHIC-STAR at $\sqrt{s_{NN}}$ = 200 GeV~\cite{STAR:2003xyj,PHENIX:2010tme}, as well as from the LHC~\cite{ATLAS:2015qwl,ALICE:2018yph}, have shown that the ratio tends to be close to 1. This deviation from ideal hydrodynamic behavior is potentially due to the effects of finite shear viscosity, non-equilibrium dynamics, flow fluctuations, and more complex flow patterns in high-energy collisions~\cite{Gombeaud:2009ye,Greco:2008fs}. 
The measurements of $v_4/v_2^2$ at high baryon density in $\sqrt{s_{NN}}$ = 2.4 GeV Au+Au collisions observed in the HADES experiment are systematically close to 0.5~\cite{HADES:2020lob,HADES:2022osk}. 
However, we know that the system's dynamics may be markedly different from those in high-energy collisions, as baryonic interactions play a dominant role in shaping the system's evolution in such a high-baryon-density regime~\cite{Mohs:2020awg,Hillmann:2019wlt}, as indicated by recent STAR flow measurements in $\sqrt{s_{NN}}$ = 3 GeV Au+Au collisions~\cite{STAR:2021yiu,Lan:2021lwk,Lan:2022rrc}.


In this study, we use the Simulating Many Accelerated Strongly Interacting Hadrons (SMASH) transport model to explore the elliptic flow $v_2$ and quadrangular flow $v_4$ in Au+Au collisions at high baryon density. 
We focus on proton flow in SMASH calculations because protons are the dominant species in regions of high baryon density. 
The primary objectives are to investigate the equation of state of baryon-rich matter and to understand the system's evolution under these conditions. 
Moreover, the HADES experiment has only published results for protons~\cite{HADES:2020lob}.
As the azimuth of the reaction plane is set to 0 in the model, the flow coefficients $v_n$ are calculated as $v_n = \langle \cos[n\phi] \rangle$, where the average is taken over all particles across all events. 
The rest of this paper is organized as follows: First, we introduce the SMASH model. 
Then, we compare $v_2$ and $v_4$, and their ratio $v_4/v_2^2$ between SMASH calculations and HADES measurements in Au+Au collisions at \mbox{$\sqrt{s_{NN}}$ = 2.4 GeV}, followed by a discussion of the time evolution and energy dependence of $v_4/v_2^2$ from SMASH. 
Finally, a summary is presented.

\section{SMASH Model}\label{smashmodel}
The SMASH model~\cite{SMASH:2016zqf} is a versatile and comprehensive hadronic transport approach designed to study the non-equilibrium dynamics of hadronic matter. 
It is particularly well suited for investigating heavy-ion collisions, where the complex interplay of numerous hadrons and their interactions must be considered. 
The model is based on the principles of the relativistic transport theory and provides a detailed description of hadronic interactions, including elastic and inelastic scatterings, resonance decays, and the propagation of hadrons through a medium. 
It offers valuable insights into the evolution of particle production, flow observables, and other aspects of heavy-ion collisions, making it a powerful tool for exploring various regimes of the QCD phase diagram.

SMASH simulates the time evolution of hadronic systems by solving the Boltzmann transport equation. 
The Hamiltonian governs the evolution of particles and is given by the sum of the single-particle energies:

\begin{equation}
    H = \sum_{i=1}^{N}\sqrt{p_{i}^{2}+m_{i}^{2}}+\sum_{i<j}V_{ij}(r_{i},r_{j})
\end{equation}
The
following equations of motion for hadrons are derived from the Hamiltonian:
\begin{equation}
    \frac{dr_{i}}{dt}=\frac{\partial H}{\partial p_{i}},
    \frac{dp_{i}}{dt}=-\frac{\partial H}{\partial r_{i}}
\end{equation}
These equations are numerically solved to track the trajectories of all hadrons in the system, where $p_i$ and $m_i$ represent the momentum and mass of the $i$-th hadron, respectively, and $V_{ij}(r_i, r_j)$ denotes the potential interaction between the $i$-th and $j$-th hadrons, depending on their positions $r_i$ and $r_j$. 
The potential, described below, incorporates various types of interactions to model the dynamics between particles. 
The specific form of the potential can vary, but typically the combined potential $V$ includes contributions from the Coulomb potential $V_C$, Yukawa potential $V_{YK}$, mean-field potential $V_{MF}$, and possibly other interaction potentials. 
These potentials are crucial for understanding the collective flow phenomena observed in such collisions.

\begin{equation}
    V_{ij}(r_i,r_j) = V_C + V_{YK} + V_{MF} + \dots
\end{equation}
The Coulomb potential accounts for the electrostatic interaction between charged particles, which is particularly important for accurately modeling the dynamics of charged hadrons. 
The Yukawa potential is used to describe the screened interaction between nucleons, reflecting the finite range of the nuclear force due to meson exchange. 
Additionally, the mean-field potential, including the Skyrme and symmetry potentials, is implemented in the calculation.
\begin{equation}
    V_{MF} = V_{Sk} + V_{Sym}
\end{equation}
The hadronic Skyrme potential, $V_{Sk}$, describes the effective interaction between nucleons in nuclear matter~\cite{Kruse:1985hy,Molitoris:1985gs}. It includes terms that depend on the local nucleon density and its gradients, which are essential for capturing the nuclear equation of state~\cite{Oliinychenko:2022uvy}:
\begin{equation}
    V_{Sk} = \alpha \left ( \frac{\rho_{B}}{\rho_{0}} \right ) + \beta \left ( \frac{\rho_{B}}{\rho_{0}} \right ) ^{\gamma}
\end{equation}
where $\rho_{B}$ is the net-baryon density; $\rho_{0}$ = 0.168 fm$^{-3}$ is the nuclear ground state density; and the parameters $\alpha$, $\beta$, and $\gamma$ define the stiffness of the nuclear equation of state. In this study, the parameters $\alpha$, $\beta$, and $\gamma$ are $-$124 MeV, 70.5 MeV, and 2, respectively, corresponding to a hard equation of state (EOS) with nuclear incompressibility of $\kappa$ = 380 MeV.
For comparison, a soft EOS is characterized by a lower nuclear incompressibility, typically \mbox{$\kappa$ = 240 MeV}~\cite{SMASH:2016zqf}. 
This choice of a hard EOS is motivated by previous studies using the UrQMD model~\cite{Hillmann:2019wlt}, which demonstrated that the hard EOS provides a better description of the HADES experimental data than the soft EOS. 
Additionally, the hard EOS also achieves better agreement with the STAR data at $\sqrt{s_{NN}}$ = 3 GeV compared to the soft EOS~\cite{STAR:2021yiu}.
These potentials are calculated after the interactions are performed, just before propagation occurs.

\section{Results}\label{results}
In the following, we will report the $v_2$ and $v_4$ of protons in the SMASH model, including both the mean-field and cascade approaches. 
The correlation between $v_2$ and $v_4$, expressed as the ratio $v_{4}/v_{2}^2$, is studied as a function of rapidity, transverse momentum, and collision centrality, respectively, along with comparisons to experimental results.

\begin{figure}[!hbt]
\centering
{\includegraphics[width=0.4\textwidth]{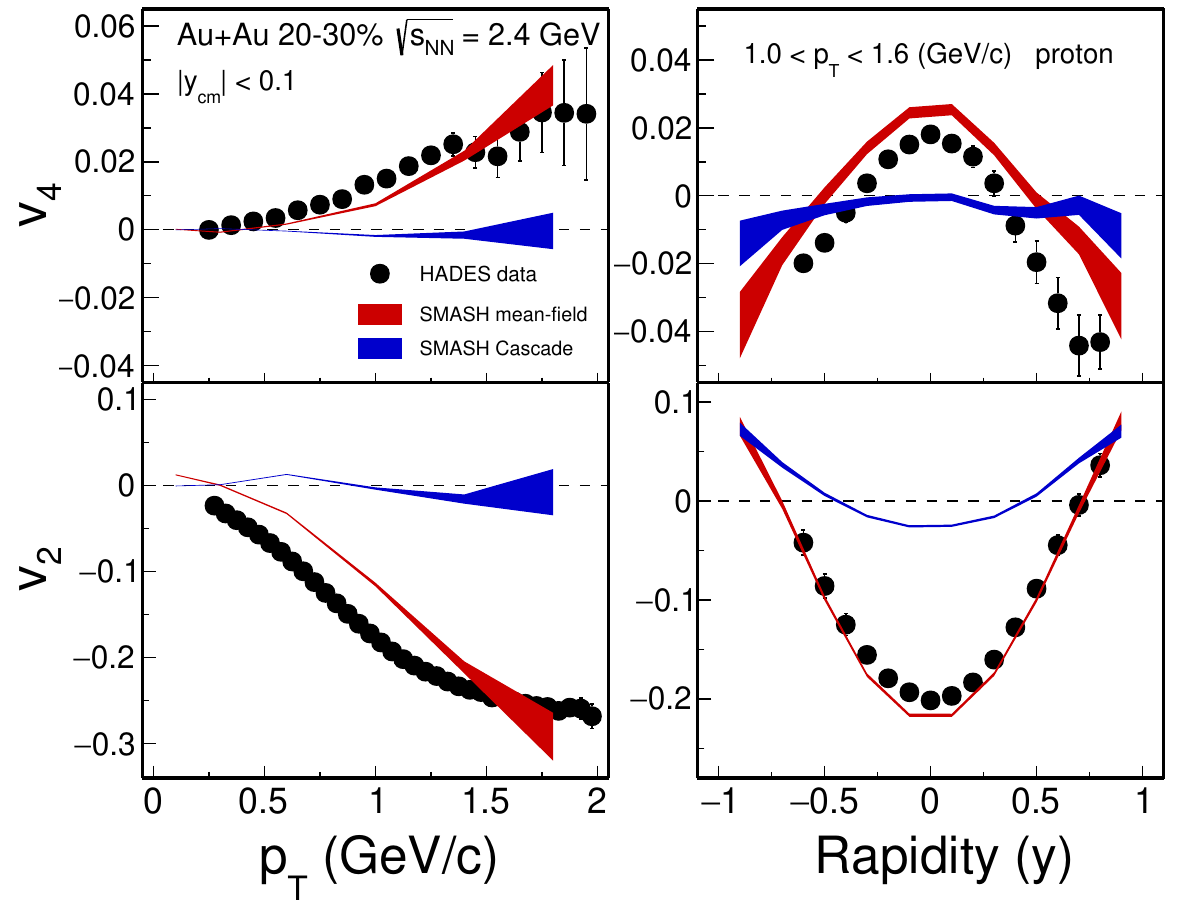}}
\caption{Comparison of $v_n$ versus $p_{T}$ and rapidity between HADES and SMASH results at 2.4 GeV.}
\label{fig:hades_com}
\end{figure}

The recent measurements of collective flow from the HADES experiment at\linebreak \mbox{$\sqrt{s_{NN}}$ = 2.4 GeV}~\cite{HADES:2020lob,HADES:2022osk} have been compared with predictions from the SMASH model, incorporating both mean-field and cascade modes. 
These model calculations were performed using the same centrality intervals and kinematic selections as the experimental data. 
Figure~\ref{fig:hades_com} illustrates the transverse momentum and rapidity dependence of elliptic flow $v_2$ and quadrangular flow $v_4$ for protons in 20--30\% centrality Au+Au collisions at $\sqrt{s_{NN}}$ = 2.4 GeV. The experimental data, shown as solid symbols, are overlaid with the SMASH predictions, where the red and blue bands correspond to the calculations from the mean-field and cascade modes, respectively.
A significant $v_4$ signal has been observed for the first time in heavy-ion collisions at such low energy by the HADES experiment. 
As shown in Figure~\ref{fig:hades_com}, the $v_4$ for protons at mid-rapidity is initially positive, but then transitions to negative values in the forward rapidity region. 
Additionally, $v_4$ increases with rising transverse momentum. In contrast to the $v_4$ behavior, the proton $v_2$ at $\sqrt{s_{NN}}$ = 2.4 GeV is negative and reaches its maximum anisotropy at mid-rapidity, increasing from mid- to forward rapidity. A strong transverse momentum dependence is also observed in the mid-rapidity region. 
The spectators, which are nucleons that do not participate in the collisions, remain primarily in the mid-rapidty region and create a `shadowing' effect. This shadowing suppresses transverse expansion in directions perpendicular to the reaction plane during the early stage of the system's evolution. Consequently, this results in a negative $v_{2}$ at mid-rapidity, as the in-plane expansion is more inhibited than the out-of plane expansion. In contrast, at forward rapidity, the influence of the spectators diminishes, as this region is less affected by their shadowing effect. This reduction in suppression allows particles at forward rapidity to undergo more transverse expansion, leading to a larger or even position $v_{2}$.

As shown in the figure, the blue bands represent the standard cascade mode calculations from the SMASH model, which fail to describe the experimental data. 
It was also found that the $v_2$ and $v_4$ signals in cascade mode are relatively weak. However, the SMASH calculations incorporating the baryonic mean-field potential, represented by the red bands, successfully reproduce both the rapidity and transverse momentum dependence of $v_2$ and $v_4$. 
The inclusion of the mean-field potential significantly enhances the signal, making it comparable to the experimental results. 
This suggests the presence of a strong baryon mean field in the high-baryon-density region. 
In comparison to the standard cascade mode, the baryon mean field generates a larger repulsive pressure, driving a stronger expansion. 
As a result, the protons are pushed out of the system earlier and with greater force.

\begin{figure}[!h]
{\includegraphics[width=0.5\textwidth]{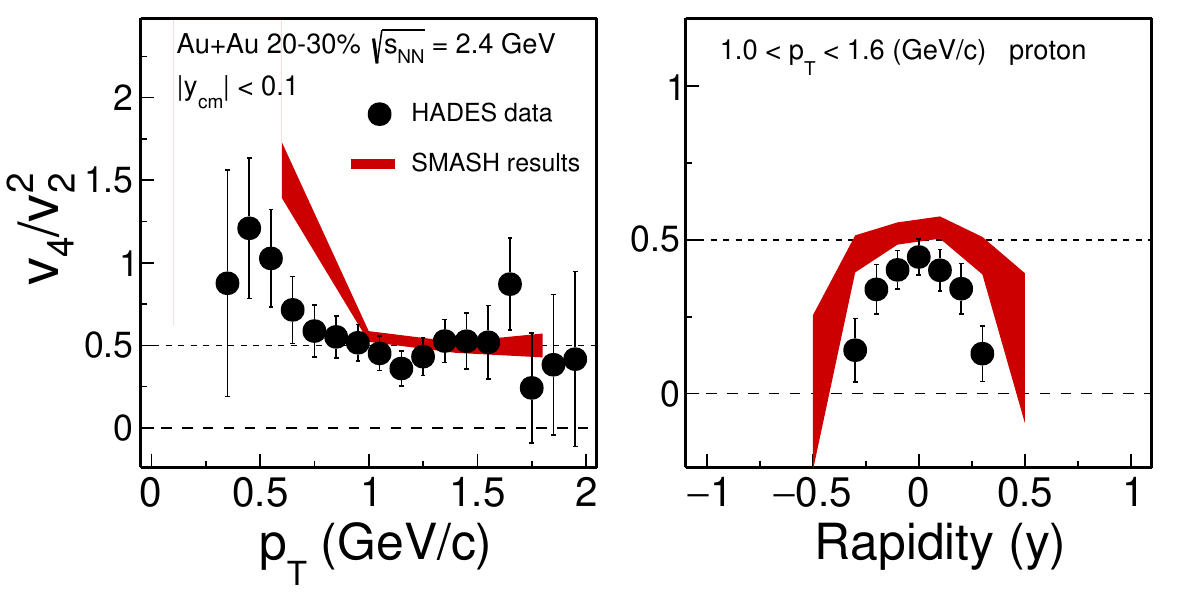}}
\caption{The ratio $v_{4}/v_{2}^{2}$ for protons as a function of $p_{T}$ and rapidity in Au+Au collisions at $\sqrt{s_{NN}}$ = 2.4 GeV from the HADES experiment and SMASH.}
\label{fig:ratio_hades}
\end{figure}

The ratio $v_{4}/v_{2}^2$ for protons in the 20--30\% centrality interval as a function of transverse momentum $p_T$ and rapidity $y$ is presented in Figure~\ref{fig:ratio_hades}. 
The HADES experimental measurements are shown by solid circles, while the SMASH calculations incorporating the mean-field potential are represented by red bands. 
As shown in the left panel of Figure~\ref{fig:ratio_hades}, the ratio $v_{4}/v_{2}^2$ measured in the HADES experiment exhibits a clear $p_{T}$-dependent behavior. The ratio decreases with increasing transverse momentum for $p_{T}<$ 1.0 GeV/$c$ and becomes approximately constant at higher $p_{T}$. The SMASH calculations, which include a mean-field potential, qualitatively describe the experimental data and reproduce a ratio of about 0.5 in the high-$p_{T}$ region ($p_{T}>$ 1.0 GeV/$c$) at mid-rapidity($|y| < $ 0.1).
The right panel shows the rapidity dependence of $v_{4}/v_{2}^2$ for protons, integrated over the high-$p_{T}$ region in the 20--30\% centrality interval. The SMASH model calculations again qualitatively reproduces the experimental data, showing a ratio close to 0.5 at mid-rapidity, which decreases significantly at forward rapidities. 
Overall, the ratio $v_{4}/v_{2}^2$ for protons demonstrates significant $p_{T}$ and rapidity dependence, as observed in both the HADES data and SMASH calculations. {Notably, the ratio calculated for high-$p_{T}$ protons \mbox{($p_{T}>$ 1.0 GeV/$c$)} at mid-rapidity \mbox{($|y|<$ 0.1)} approaches a value of approximately 0.5, surprisingly aligning with expectations based on hydrodynamic behavior~\cite{Borghini:2005kd}.} 
 However, this result should be interpreted as a qualitative agreement with the experimental data rather than a direct evidence of hydrodynamic behavior. 

\begin{figure}[!h]
{\includegraphics[width=0.5\textwidth]{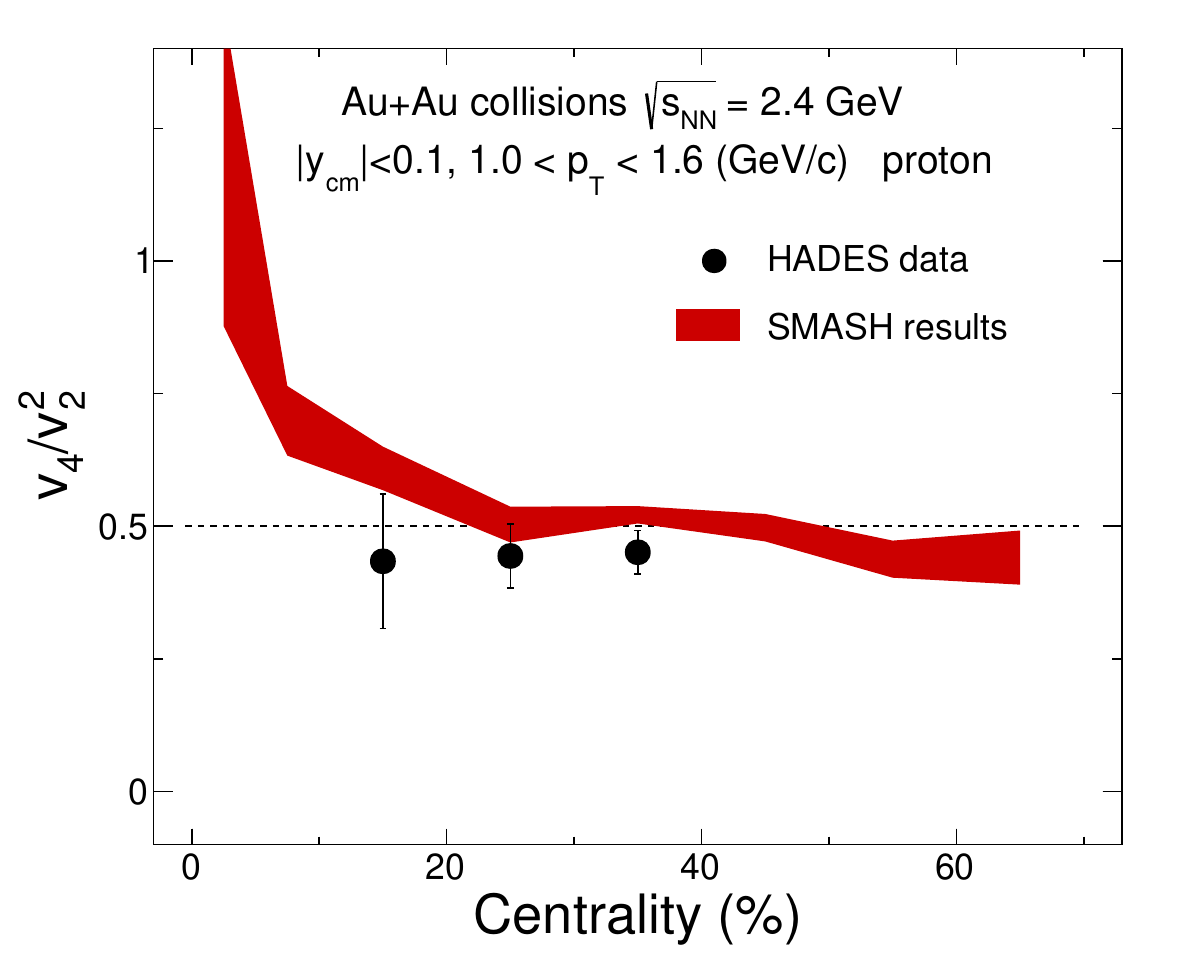}}
\caption{The ratio $v_{4}/v_{2}^{2}$ for protons as a function of centrality from SMASH Au+Au collisions with the mean-field potential. 
The solid black circle represents results from the HADES experiment in $\sqrt{s_{NN}}$ = 2.4 GeV collisions.}
\label{fig:ratio_cent}
\end{figure}

According to Figure~\ref{fig:ratio_hades}, there is no significant rapidity or $p_T$ dependence of $v_{4}/v_{2}^{2}$ within the $p_T$ window ($p_T$ $\geq$ 1.0 GeV/$c$) and the rapidity window ($|y|$ $\leq$ 0.1), so we integrate over these variables.
Figure~\ref{fig:ratio_cent} shows the centrality dependence of the $v_{4}/v_{2}^{2}$ ratio for protons in Au+Au collisions at $\sqrt{s_{NN}}$ = 2.4 GeV, as measured by the HADES experiment~\cite{HADES:2020lob} and calculated using the SMASH model with the mean-field potential.
The SMASH model successfully reproduces the approximately flat $v_{4}/v_{2}^{2}$ ratio in 10--40\% central collisions, where the ratio is around 0.5.
Notably, the $v_{4}/v_{2}^{2}$ ratio exceeds 0.5 in the most-central collisions compared to mid-central collisions, consistent with observations at \mbox{$\sqrt{s_{NN}}$ = 200 GeV}~\cite{PHENIX:2010tme}. 
This increase is likely due to contributions from eccentricity fluctuations and flow fluctuations~\cite{Gombeaud:2009ye,Greco:2008fs}.

\begin{figure}[!h]
{\includegraphics[width=0.5\textwidth]{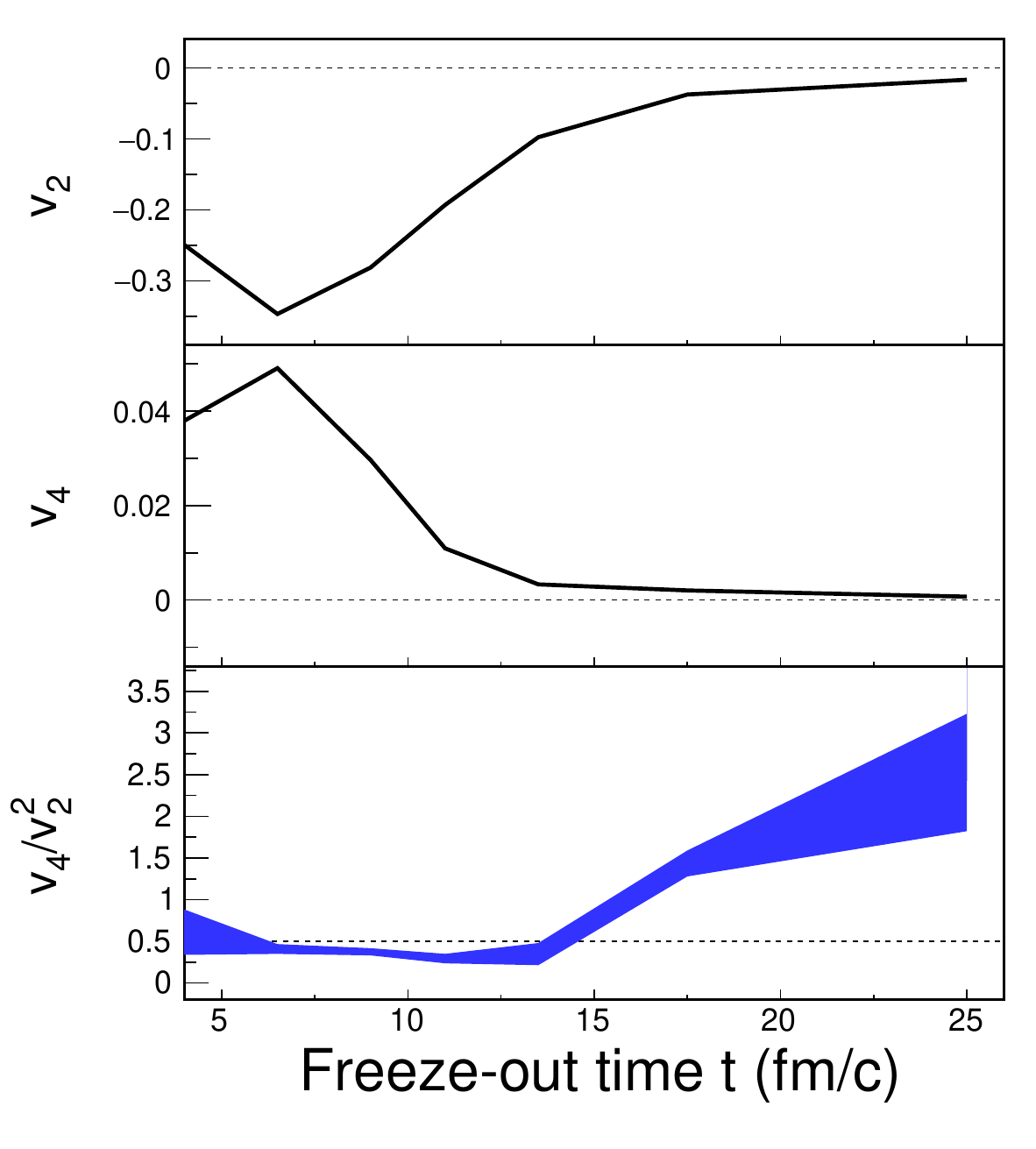}}
\caption{$v_{2}$, $v_{4}$, and the ratio $v_{4}/v_{2}^{2}$ for protons as a function of kinetic freeze-out time.}
\label{fig:ratio_dt}
\end{figure}

Next, we examine the time evolution of elliptic flow and quadrangular flow to investigate the transformation of nuclear matter. 
The freeze-out time on the x-axis of Figure~\ref{fig:ratio_dt} refers to the kinetic freeze-out time of protons, within $t$ $\pm$ 1 fm/$c$.
The first and second panels of Figure~\ref{fig:ratio_dt} show $v_2$ and $v_4$ of protons as a function of the kinetic freeze-out time in Au+Au 20--30\% collisions at $\sqrt{s_{NN}}$ = 2.4 GeV, using the SMASH model with the mean-field potential.
At the early stage of expansion (freeze-out time $t$ $\lesssim$ 10 fm/$c$), $v_2$ exhibits well-defined negative values, while $v_4$ shows prominent positive values, reflecting the initial collision geometry with large eccentricity. 
Note that a negative $v_2$ implies emission out-of-plane due to the strong spectator shadowing effect~\cite{E895:1999ldn, STAR:2021yiu}. 
As time progresses \mbox{($t \gtrsim$ 10 fm/$c$)}, the magnitudes of $v_2$ and $v_4$ gradually decrease and approach zero. 
This indicates that the initial anisotropic pressure gradients of the dense bulk vanish, and the system becomes more isotropic as the evolution progresses.

The bottom panel of Figure~\ref{fig:ratio_dt} depicts the evolution of the $v_{4}/v_{2}^{2}$ ratio as a function of kinetic freeze-out time, based on the $v_{2}$ and $v_{4}$ values shown in the top two panels.
The results indicate that the $v_{4}/v_{2}^{2}$ ratio remains approximately constant at around 0.5 during the early stages of the system's expansion ($t \lesssim$ 15 fm/$c$) and gradually increases as the system evolves in the later stage.

\begin{figure}[!h]
{\includegraphics[width=0.5\textwidth]{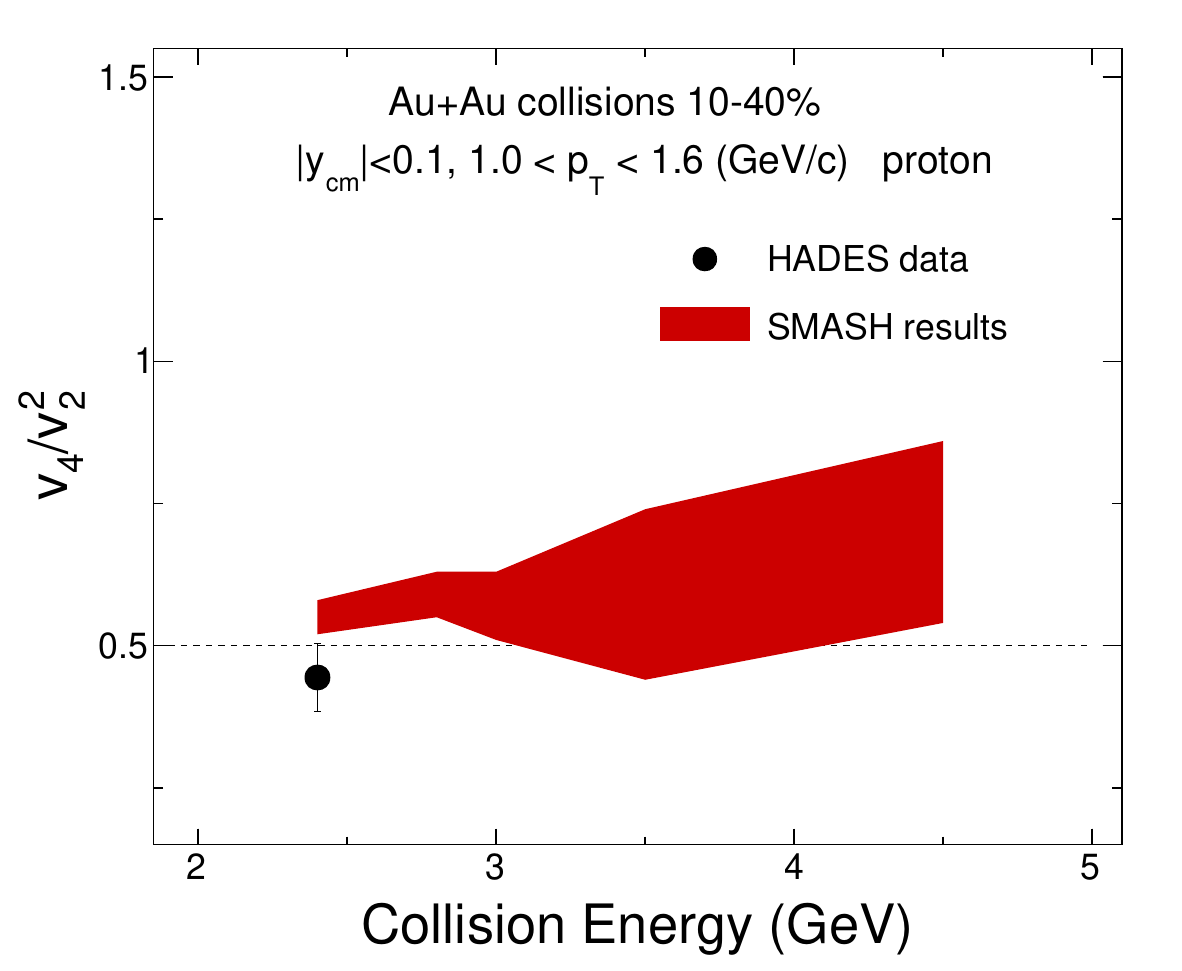}}
\caption{The ratio $v_{4}/v_{2}^{2}$ for protons as a function of collision energy in 10--40\% centrality Au+Au collisions, calculated using the SMASH model with a mean-field potential at $\sqrt{s_{NN}}$ = 2.4, 2.8, 3.0, 3.5, and 4.5 GeV. The black solid circle represents results from the HADES experiment for Au+Au collisions at $\sqrt{s_{NN}}$ = 2.4 GeV and 20--30\% collisions.}
\label{fig:ratio_ene}
\end{figure}

Figure \ref{fig:ratio_ene} shows the energy dependence of the ratio $v_{4}/v_{2}^{2}$ for protons in Au+Au collisions at 10--40\% centrality, calculated using the SMASH model with the mean-field potential at $\sqrt{s_{NN}}$ = 2.4, 2.8, 3.0, 3.5, and 4.5 GeV.
The black solid circles represent data from the HADES experiment for Au+Au collisions at $\sqrt{s_{NN}}$ = 2.4 GeV and 20--30\% centrality.
In the model calculations, the transverse momentum range is limited to 1.0 $< p_{T} <$ 1.6 (GeV/$c$), and rapidity is constrained to $|y_{cm}| < 0.1$, matching the selection criteria of the HADES experiment and focusing on mid-rapidity protons.
This study targets the high-baryon-density region, with collision energies ranging from 2.4 to 4.5 GeV, covering the energy range of upcoming facilities such as FAIR, NICA, and HIAF. 
The SMASH calculations, shown by the red band in the figure, are qualitatively consistent with a value of 0.5 within uncertainties.
The $v_4$ value is very small when $\sqrt{s_{NN}} >$ 3 GeV, reaching $v_4 \approx 0.001$ at 4.5 GeV, leading to significant uncertainties due to error propagation. Based on these transport model calculations, the ratio $v_{4}/v_{2}^{2}$ remains consistent with 0.5 within uncertainties, with no significant energy dependence observed in the high-baryon-density region.

Regarding the results shown above, the pure transport model calculations yield outcomes that are consistent with the HADES data for this observable. This indicates that, in low-energy nucleon collisions, the system's response to the initial geometry, as modeled through a transport approach, is reasonably captured in the final-state observable. However, further investigations are needed to fully understand the underlying dynamics at such low collision energies.

\section{Summary}\label{summary}
In summary, we have investigated the elliptic and quadrangular flow of protons in Au+Au collisions at $\sqrt{s_{NN}}$ = 2.4 GeV using the hadronic transport model SMASH. 
By incorporating the mean-field potential, our results qualitatively reproduce the measured elliptic and quadrangular flow, as well as the ratio $v_{4}/v_{2}^{2}$ observed in the HADES experiment. 
This suggests that baryonic interactions play a crucial role in shaping anisotropic flow in the high-baryon-density region.

We further examined the centrality dependence of the $v_{4}/v_{2}^{2}$ ratio for protons within the SMASH mean-field framework, finding a value around 0.5 in mid-central and peripheral collisions, consistent with the HADES experiment measurements. 
Additionally, the time evolution of $v_2$, $v_{4}$, and their ratio $v_{4}/v_{2}^{2}$ shows that the ratio remains close to 0.5 during flow development, particularly when the freeze-out time is less than 15 fm/$c$.

Lastly, we explored the energy dependence of $v_{4}/v_{2}^{2}$ in the high-baryon-density region. 
Our results suggest that the ratio remains constant and close to 0.5 in this regime, consistent with the observation from the HADES experiment. 
Further theoretical and experimental studies are needed to better understand the underlying mechanisms at low collision energies.
These findings provide valuable insights into the anisotropic flow development at high baryon density and will help guide future experiments from upcoming facilities such as FAIR, NICA, and HIAF.

\section{acknowledgments}


This work was supported by the Doctoral Scientific Research Foundation of Pingdingshan University (PXY-BSQD-2023016),
the National Key Research and Development Program of China under Contract No. 2022YFA1604900, the National Natural Science Foundation of China (NSFC) under contract No. 12175084.

\newpage

\end{document}